# The Reactivity of MgB$_2$ with Common Substrate and Electronic Materials

T. He[1], John M. Rowell[2], and R.J. Cava[1]

[1]*Department of Chemistry and Princeton Materials Institute*
*Princeton University, Princeton NJ*
[2]*Materials Research Institute, Northwestern University, Evanston IL*

**Abstract**

The reactivity of MgB$_2$ with powdered forms of common substrate and electronic materials is reported. Reaction temperatures between 600°C and 800°C, encompassing the range commonly employed in thin-film fabrication, were studied. The materials tested for reactivity were ZrO$_2$, yttria stabilized zirconia (YSZ), MgO, Al$_2$O$_3$, SiO$_2$, SrTiO$_3$, TiN, TaN, AlN, Si, and SiC. At 600°C, MgB$_2$ reacted only with SiO$_2$ and Si. At 800°C, however, reactions were observed for MgB$_2$ with Al$_2$O$_3$, SiO$_2$, Si, SiC, and SrTiO$_3$. The T$_c$ of MgB$_2$ decreased in the reactions with SiC and Al$_2$O$_3$, $MgB_2$, with a Tc of 39K (1), may offer higher operating temperatures and device speeds than today's Nb-based technologies for potential applications in thin-film electronic devices, and may have a simpler multilayer film fabrication process than high temperature superconductors. An extensive review of the progress in thin-film fabrication of $MgB_2$ has been presented (2). For electronics applications, it is desirable that films with $T_c$s close to 39K be made by a "single-step in-situ" process, in which $MgB_2$ is formed directly on substrates. The majority of films to date have been made by a two-step process, in which a precursor film of either B or Mg + B is annealed, typically either in Mg vapor at 900°C, or in inert gas at about 600°C, though some single-step processes have also been reported. In all cases, the reactivity of $MgB_2$ with substrate materials or insulating or metallic layers in multi-layer circuits is an important factor in determining both the conditions of film fabrication and the operating characteristics of the resulting device. Here we describe the results of experiments in which fine powders of different materials were reacted with Mg + B in the 600 - 800°C temperature range relevant to device fabrication. It was found that $MgB_2$ is inert toward many common electronic and substrate materials at 600°C (with the notable exceptions of $SiO_2$ and Si), but is highly reactive toward some of the most common substrate materials (i.e. $Al_2O_3$ and $SrTiO_3$) by 800°C, suggesting their limited usefulness if high fabrication temperatures must ultimately be used during film deposition.

Polycrystalline samples of $MgB_2$ plus various materials commonly used in thin film devices were made by solid state reaction. Starting materials were bright Mg flakes, sub-micron amorphous B powder, and pre-reacted electronic materials obtained commercially. These materials were in fine powder form, to enhance reactivity, and were employed as received.

They were $ZrO_2$, yttria stabilized zirconia (YSZ), MgO, $Al_2O_3$, $SrTiO_3$, TiN, TaN, AlN, $SiO_2$ (crystalline), Si, and SiC. Elemental Mg + B were employed in the reactions, rather than pre-formed $MgB_2$, to better model film fabrication processes. The ratio of $MgB_2$ to reactant material was set to a 9:1 mole ratio. Starting materials were mixed thoroughly in quarter-gram batches, and pressed into pellets to enhance reaction rates. The pellets were placed on Ta foil, which was in turn placed on a dense $Al_2O_3$ boat, and heated in a quartz tube furnace under a mixed gas of 95% Ar 5% $H_2$. Each set of pellets was heated for 24 hours at one of three temperatures: 600°C, 700°C, or 800°C. Due to Mg volatility and the relatively long heating time, 20% excess Mg was employed in these reactions. To get a clear test of the reactivity at 800°C, where the volatility of Mg in open systems at long heating times is significant (3), those tests were repeated by sealing the starting materials in stoichiometric proportions in Ta tubes back-filled with Ar. Those tubes were then sealed in evacuated quartz tubes, and heated in a box furnace at 800°C for 10 hours. Reported results for stability at 800°C are taken from those samples.

The phases present after annealing were determined by powder X-ray diffraction (XRD) with Cu K$\alpha$ radiation at room temperature. The results of all the reactivity studies are summarized in table 1 and figures 1-3. $MgB_2$ was found to be inert with respect to $ZrO_2$, YSZ, MgO, TiN and AlN up to 700°C (only $MgB_2$ plus reactant compound starting material was observed in the XRD patterns). In an earlier study (4) it was shown that bulk $MgB_2$ prepared at temperatures as low as 550°C displayed a $T_c$ equivalent to the bulk value of 38-39K, and the same $T_c$ was observed in the above cases. $MgB_2$ was found to be inert with respect to $Al_2O_3$ only at 600°C. At higher temperatures, MgO was formed as a reaction product in the Mg + B

+ $Al_2O_3$ system, and the diffraction pattern for the $MgB_2$ showed a significant change in the unit cell parameters, consistent with the incorporation of significant quantities of Al into the $MgB_2$ structure. Al incorporation in $MgB_2$ is known to decrease $T_c$ (5). $MgB_2$ was stable with respect to $SrTiO_3$ up to 700°C. However, $SrB_6$, $MgB_4$ and several peaks due to unknown phases were found when $MgB_2$ and $SrTiO_3$ were reacted at 800°C, indicating that $SrTiO_3$ and $MgB_2$ are not compatible at high temperatures. The chemical compatibility problems of $MgB_2$ with $Al_2O_3$ and $SrTiO_3$, which are among the most commonly used substrates, are therefore serious. Clearly there were reactions between $MgB_2$ and those two oxides, and caution should be taken when fabricating thin films.

Figures 1 and 2 show the XRD data for the reaction of $MgB_2$ with six oxide substrate materials ($ZrO_2$ and YSZ are often employed as oxide buffer layers) at 800°C. No reaction occurred when the substrates were $ZrO_2$ and MgO, suggesting them to be potentially good substrates for making thin films. A small amount of MgO was formed in the sample of $MgB_2$ reacted with YSZ, but it is also a good candidate substrate. $SiO_2$ is found to be highly reactive with $MgB_2$ even at 600°C. This result implies that oxidized silica wafers should not be used for $MgB_2$ film devices unless a buffer layer, such as MgO, YSZ, or $ZrO_2$ is first deposited on the $SiO_2$. Figure 3 shows the XRD data for the reaction of $MgB_2$ with four non-oxide materials, TiN, TaN, AlN and SiC. $MgB_2$ as found to be fully stable with respect to the nitrides TaN, TiN and AlN up to 800°C. The nitrides are not common substrates but are employed in thin film devices such as bolometers. In the case of SiC, however, the peaks of $MgB_2$ were shifted, indicating that some reaction had occurred. $MgB_2$ was found to be highly reactive with Si at temperatures of 600°C and higher (fig. 2), suggesting that these materials are chemically incompatible. Study of the compatibility of $MgB_2$ with $Mg_2Si$ (a cubic semiconductor), the decomposition product formed during the reaction, would be of interest to see whether a buffer layer of $Mg_2Si$ might passivate the surface of Si in multilayer devices with $MgB_2$.

To determine the effect of the reactions on the superconducting transition temperature, the low temperature magnetization was measured for all the samples made in Ta tubes at 800°C. Samples were measured in the form of loose powders in a Quantum Design PPMS magnetometer with an applied DC field of 15 Oe. The zero-field cooled DC magnetic susceptibility data for the $MgB_2$ reacted with six oxide materials are presented in Fig. 4. Except for the case of $Al_2O_3$, the $MgB_2$ present in reactions with the other four oxide substrate materials had a superconducting $T_c$ of about 38 K, which is just the $T_c$ of pure $MgB_2$ (1). $SrTiO_3$ and $SiO_2$, although reacting with $MgB_2$ and causing partial decomposition, had no effect on the $T_c$ of any $MgB_2$ present. The sample reacted with $Al_2O_3$, however, showed a broad superconducting transition with a suppression of $T_c$ of about 3 K, consistent with the partial substitution of Al for Mg in the $MgB_2$. The zero-field cooled DC magnetic susceptibility data for $MgB_2$ reacted with five non-oxide materials are presented in Fig. 5. $MgB_2$ heated in the presence of all three nitrides, TiN, TaN and AlN, had a superconducting transition at 38 K, consistent with the lack of reaction seen in the XRD patterns. Reaction with Si did not decrease the $T_c$ of any $MgB_2$ present. SiC, however, suppressed $T_c$. There have been no reports on SiC doping of $MgB_2$. Carbon, however, has been doped into $MgB_2$, and was found to decrease $T_c$ (6), which may be the origin of the degradation observed here.

Surprisingly, $MgB_2$ has been found to be inert with respect to many electronic materials at 600°C, suggesting that the chemical compatibility of $MgB_2$ with other electronic materials in low temperature fabrication processes will not generally be an important factor in determining device performance. Important exceptions to this overall character are $SiO_2$ and Si, where reactivity is observed even at 600°C, and $Al_2O_3$, where reaction was observed at 700°C. At higher temperatures, $MgB_2$ is also quite reactive with $SrTiO_3$, but is stable against some oxides and all the nitrides tested, with MgO, $ZrO_2$ and YSZ particularly promising as potential substrates and buffer layers. $MgB_2$ shows much more overall reactivity at 800°C, however. This increased reactivity of $MgB_2$ at 800°C illustrates the importance of pursuing efforts to reduce $MgB_2$ deposition temperatures to 600°C or lower for thin film device fabrication.

**Acknowledgements**

This work was performed with the support of the Office of Naval Research, grant N00014-01-1-0920.

Table 1: Reactivity of $MgB_2$ with various electronic materials.

| Electronic Material | 600°C Anneal | 800°C Anneal |
|---|---|---|
| $ZrO_2$ | No Reaction | No Reaction |
| $YSZ^1$ | No Reaction | $MgB_2$, small amount of MgO |
| MgO | No Reaction | No Reaction |
| $Al_2O_3$ | No Reaction | $MgB_2$ with altered cell size, MgO, unknown |
| $SiO_2$ | $MgB_2$, MgO, Si | $MgB_2$, $MgB_4$ MgO, $Mg_2Si$, Si |
| $SrTiO_3$ | No Reaction | $MgB_2$, $SrTiO_3$, MgO $SrB_6$, $TiB_2$ |
| Si | $MgB_2$, $Mg_2Si$ | $MgB_2$, $Mg_2Si$, $MgB_4$ |
| TiN | No Reaction | No Reaction |
| $TaN^2$ | No Reaction | No Reaction |
| AlN | No Reaction | No Reaction |
| SiC | No Reaction | $MgB_2$ with altered cell size |

[1] $ZrO_2$ is present in the YSZ before reaction.
[2] $TaN_{0.8}$ is present in the TaN before reaction.

**Figure Captions**

Fig. 1  Powder X-ray diffraction patterns (Cu Kα radiation) of $MgB_2$ reacted for 10 hours at 800°C with finely divided powders of MgO, YSZ (yttria stabilized zirconia), and $ZrO_2$ showing good chemical compatibility. The YSZ powder contained $ZrO_2$ in its pre-reacted state.

Fig. 2  Powder X-ray diffraction patterns (Cu Kα radiation) of $MgB_2$ reacted for 10 hours at 800°C with finely divided powders of $SrTiO_3$, $Al_2O_3$, crystalline $SiO_2$, and Si, showing extensive chemical reaction.

Fig. 3  Powder X-ray diffraction patterns (Cu Kα radiation) of $MgB_2$ reacted for 10 hours at 800°C with finely divided powders of SiC, AlN, TaN, and TiN. The $TaN_{0.8}$ was present in the "TaN" in its pre-reacted state.

Fig 4. Magnetic characterization of the superconducting transitions for samples of $MgB_2$ reacted with the oxides $Al_2O_3$, $SiO_2$, MgO, $ZrO_2$, YSZ and $SrTiO_3$ at 800°C.

Fig 5. Magnetic characterization of the superconducting transitions for samples of $MgB_2$ reacted with Si, SiC, and the nitrides AlN, TaN and TiN at 800°C.

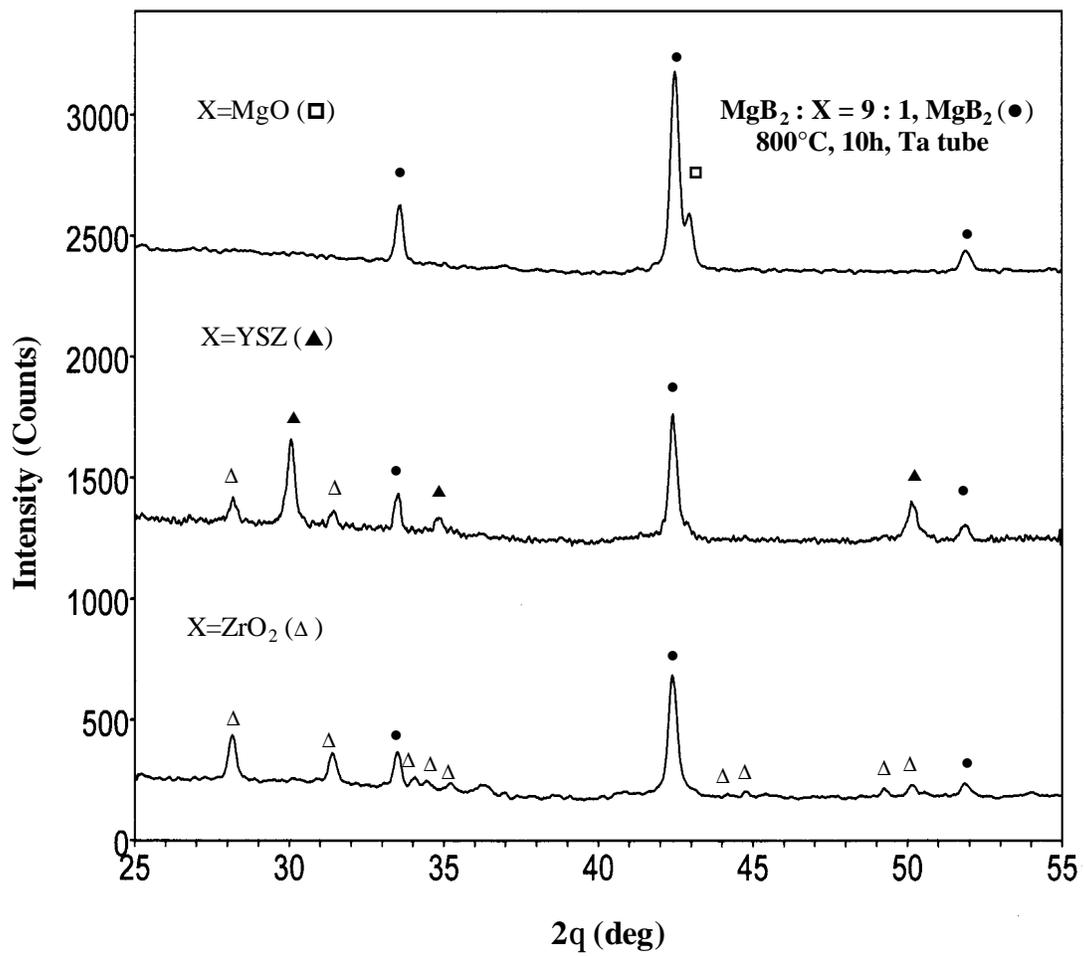

Figure 1

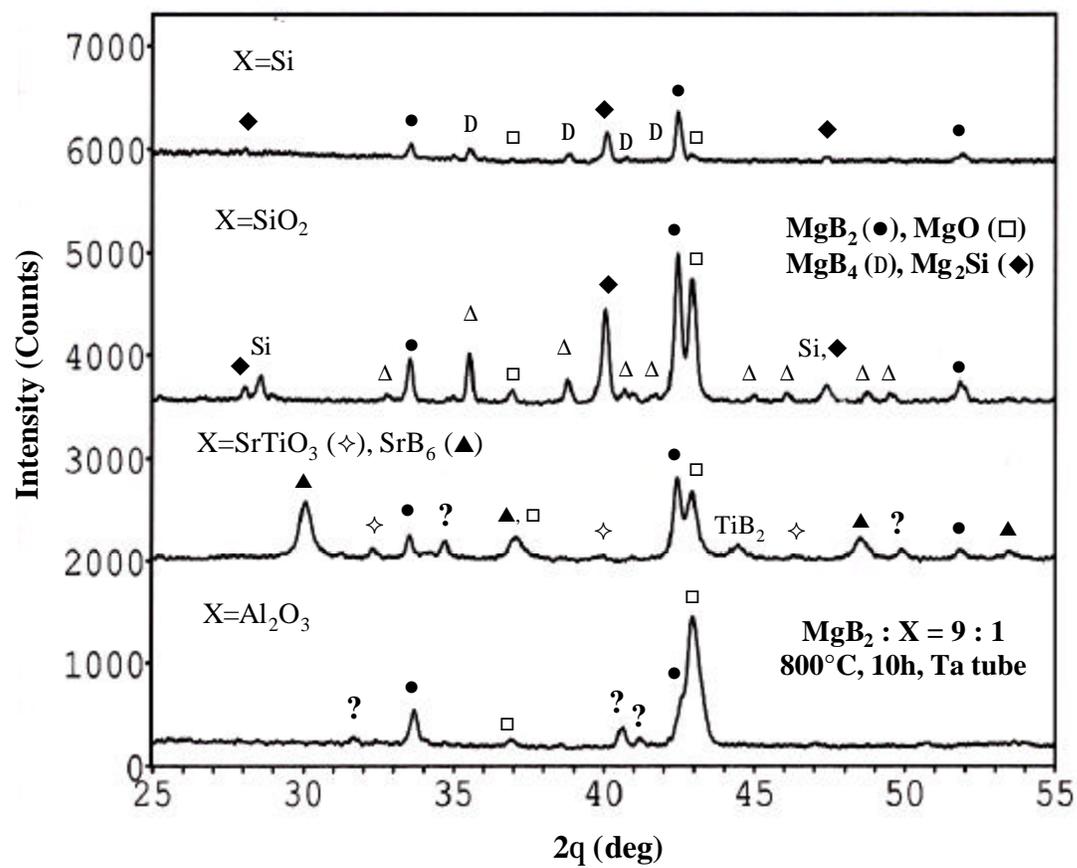

Figure 2

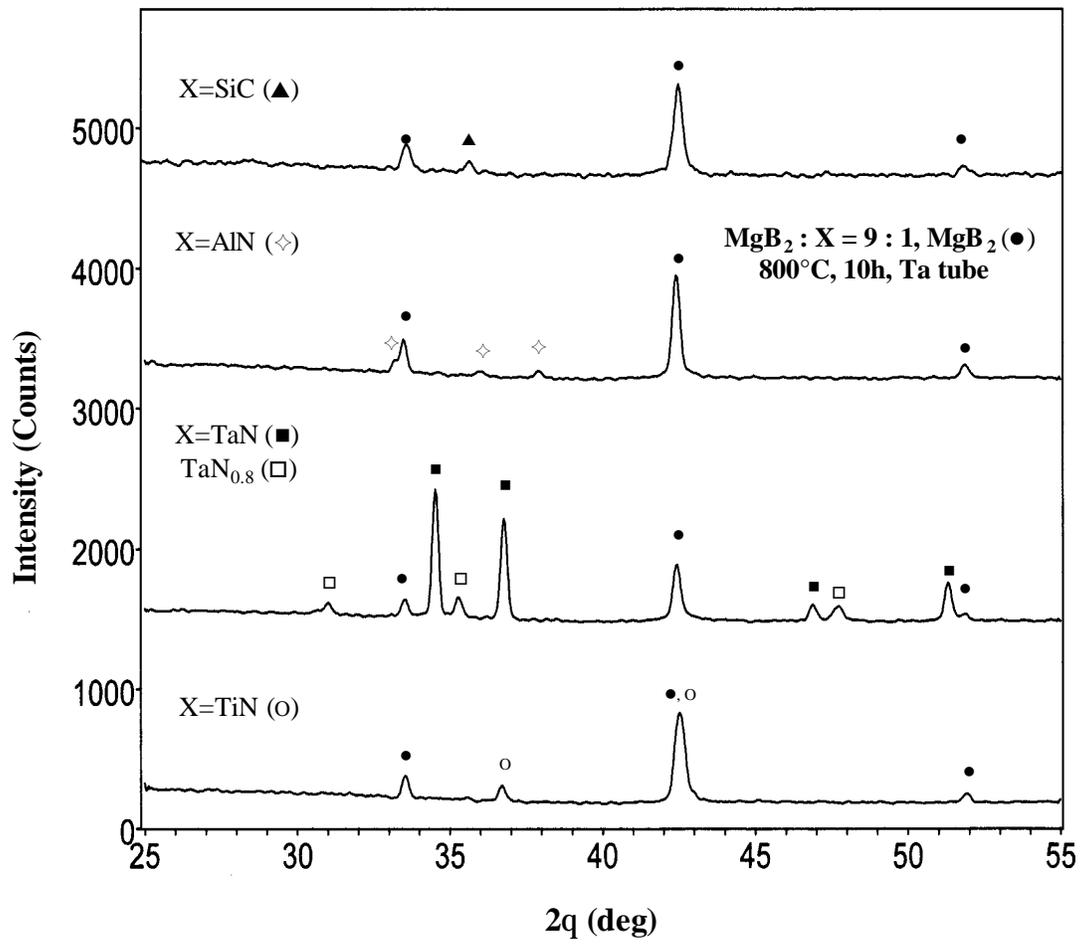

Figure 3

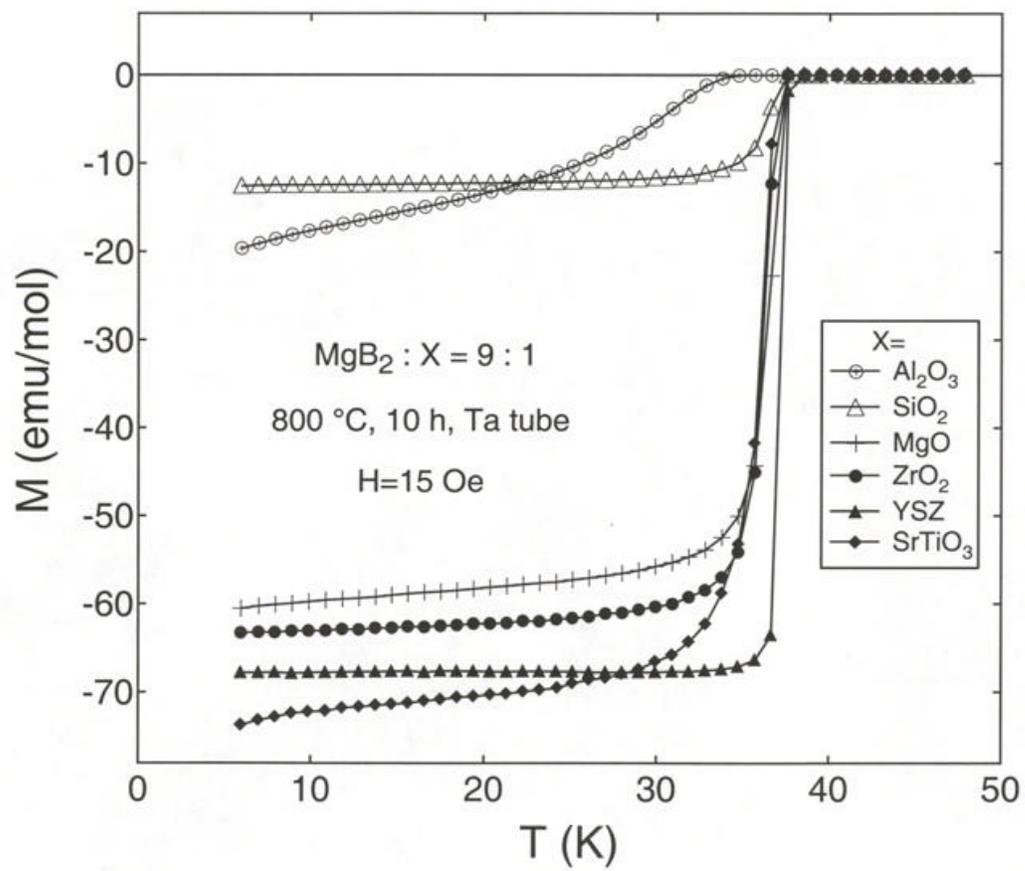

Figure 4

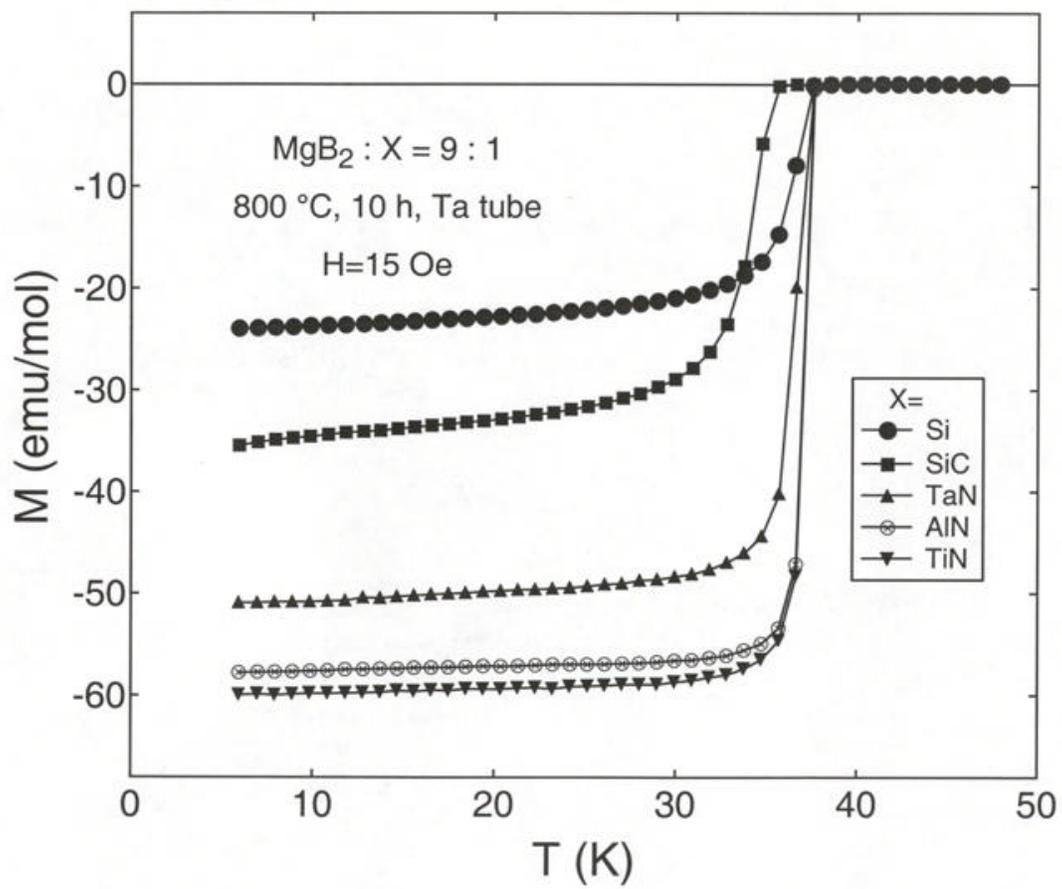

Figure 5